\documentclass{article}
\usepackage{graphicx}
\usepackage{hyperref}
\usepackage[]{natbib}  
\def\BibTeX{{\rm B\kern-.05em{\sc i\kern-.025em b}\kern-.08em
    T\kern-.1667em\lower.7ex\hbox{E}\kern-.125emX}}
\bibpunct{(}{)}{;}{a}{}{,}  

\begin{document}


\title{Dome C site testing: long term statistics of integrated optical turbulence parameters at ground level }

\author{E. Aristidi,and the Astroconcordia team}
\date{Laboratoire Lagrange, Parc Valrose, F 06108 Nice Cedex 2}


\maketitle


\begin{abstract}
We present long term site testing statistics obtained at Dome C, Antarctica with various experiments deployed within the Astroconcordia programme since 2003. We give values of integrated turbulence parameters in the visible at ground level and above the surface layer, vertical profiles of the structure constant $C_n^2$ and a statistics of the thickness of the turbulent surface layer. 
\end{abstract}


\section{Introduction}
The AstroConcordia program has been up to now, dedicated to the qualification of the site of Dome C, Antarctica, for astronomical purposes. After almost 10 years of operation since the first results~\citep{2003AA...406L..19A} including 8 winterovers, we could measure long terms characteristics of the turbulent atmosphere and extract important parameters of the so-called ``boundary layer" that we now understand quite well (global and temporal statistics). With the huge volume of data collected throughout the years, Dome C is probably among the most extensively characterized sites in the world, at least from the optical turbulence point of view.
In this contribution we present the results of 8 years of turbulence monitoring. We give statistics of the seeing, the isoplanatic angle, the coherence time and the outer scale during the polar winter, at ground level as well as above the turbulent surface layer.  We also give statistics of the thickness of the surface layer as estimated by various instruments.

\section{The instruments}
The instruments that were deployed on the site aim at characterizing the statistical properties of the optical turbulence. They can be split into two families: the ``profilers'' and the ``integrators''. The latter is composed of telescope-based instruments which observe a bright star in the visible wavelengths and perform continuous measurements of the seeing, the isoplanatic angle and the outer scale. These instruments named DIMM, Thetameter and GSM are described in previous papers (\citealt{2005AA...444..651A}, 2009, \citealt{2008AA...491..917Z}). Several DIMM were placed at different elevations to measure the seeing inside the turbulent surface layer discovered during the first winterover in 2005 \citep{2006PASP..118..344A}.

The other family includes all instruments capable of measuring vertical profiles of the refractive index structure constant $C_n^2(h)$ as a function of the altitude $h$. The balloon-borne microthermal sensors used in 2005 \citep{2008PASP..120..203T} and the sonic anemometers on the 45~m high mast~\citep{2008SPIE.7012E.147T} belong to this group and perform in-situ measurements (in the whole atmosphere for the balloons, inside the surface layer for the sonics). The Single Star Scidar~\citep{2009AA...500.1271V} and the Lunar Limb Profiler use optical techniques to access the turbulence profile though the observation of a bright star (scidar) or the lunar/solar limp (LLP). The latter was installed at Dome C in 2011 and the first winter data are under processing at the moment of writing this paper. Note that integrated parameters (including the coherence time of turbulence) can also be derived from some of these profilers (Balloons, SSS, LLP).

The period of operation of each instrument is summarized in Table~\ref{aristidi:table1}.

\section{Results}

\subsection{Statistics of integrated parameters}
\begin{figure}[ht!]
 \centering
 \includegraphics[width=0.5\textwidth,clip]{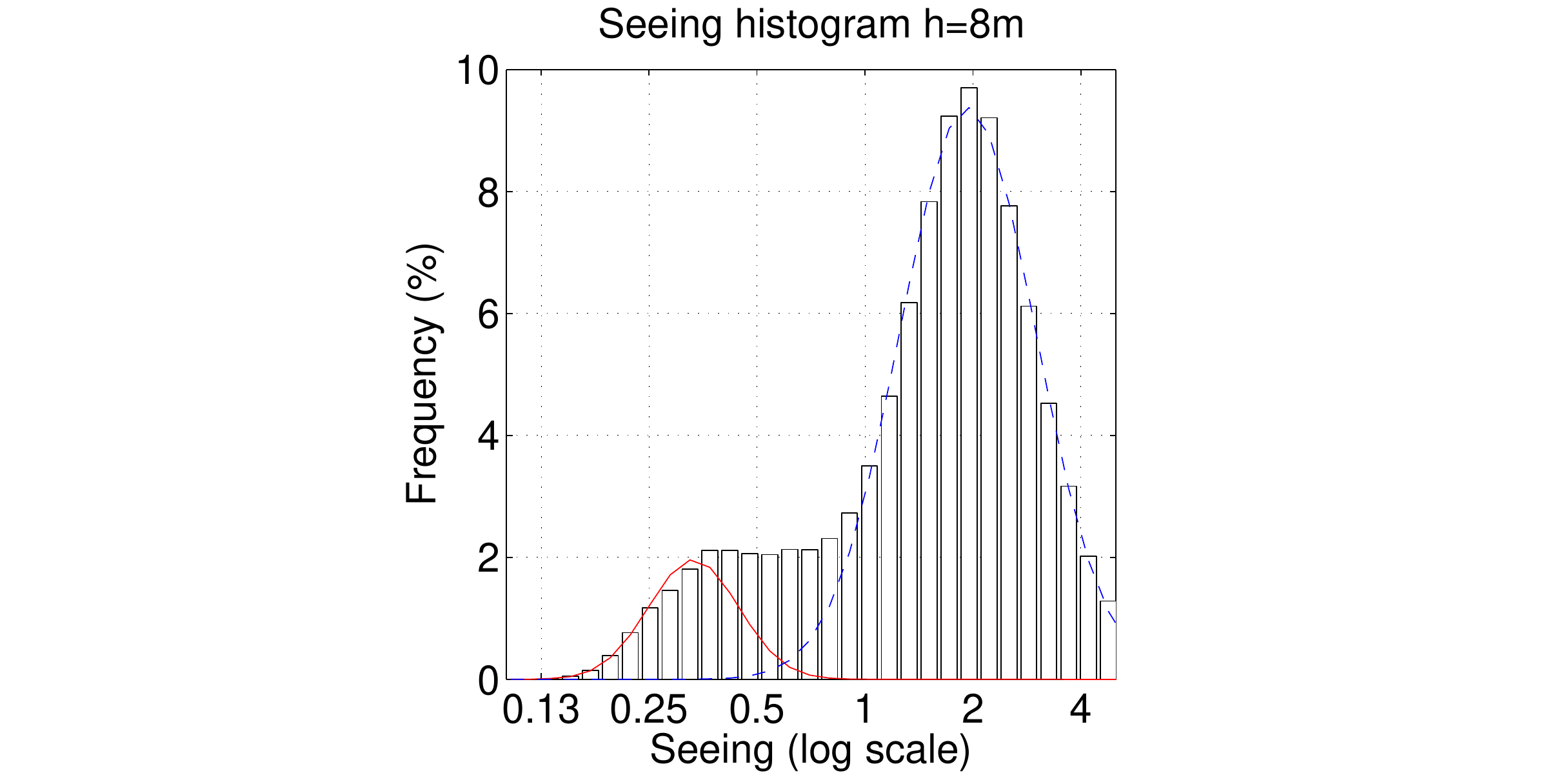}%
 \includegraphics[width=0.5\textwidth,clip]{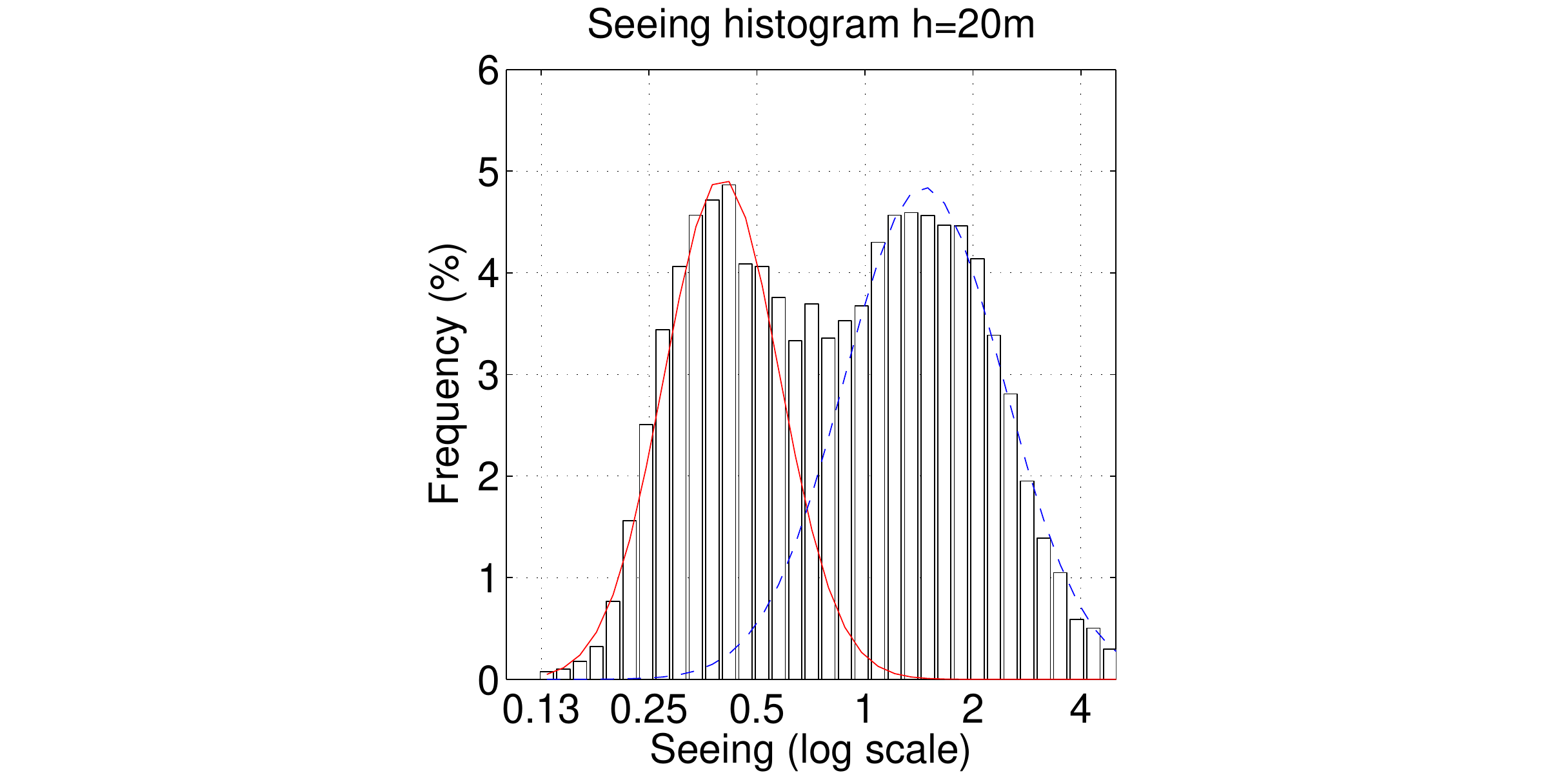}      
  \caption{{\bf Left:} Seeing histogram at an elevation of $h=8m$ for the period April to September. {\bf Right:} Idem for $h=20m$ from available data in 2005 and 2012. Red (solid) and blue (dashed) lines are Gaussian fit of the two bumps. They give the fraction of time spent by the DIMM inside (blue line) or outside (red line) the SL. At $h=20m$ the DIMM is in the free atmosphere 45\% of the time (12\% at $h=8m$).}
  \label{aristidi:fig1}
\end{figure}

A complete statistics of integrated parameters in winter (period April to September) obtained with all the instruments is presented in table~\ref{aristidi:table2}. Values are given for an altitude of 8~m above the snow. The seeing is actually poor because of the presence of a very turbulent surface layer (SL), but it was discovered~\citep{2009AA...499..955A} that the upper edge of this SL is very sharp. The seeing distribution splits into two regimes corresponding to the DIMM being inside or outside the SL, giving histograms composed of two bumps (Fig.~\ref{aristidi:fig1}). The area under these bumps depends on the altitude $h$ above the ground: at $h=20$~m, the DIMM spent 45\% of the time above the SL. Statistics of the integrated parameters above the SL could be derived from profiler data (Balloons and SSS) as well as combined analysis of measurements from the 3 DIMM~\citep{2009AA...499..955A}. They are given in Table~\ref{aristidi:table3}. 

As noticed by~\citet{2010PASP..122.1122B} for the case of Dome A which present similar properties, this situation is very peculiar. We are now convinced that site characterization in Antarctica can be performed by (i) measuring the properties of the SL (in particular its thickness) and (ii) measuring the free atmosphere turbulence parameters.

\subsection{Thickness of the surface layer}
The presence of a very turbulent surface layer in the first tens of meters above the ice of the plateau was discovered with the first balloon launches in March 2005~\citep{2006PASP..118..344A}. Several experiments were deployed to monitor this SL by in-situ measurements. The first one is a set of microthermal sensors (same sensors used on the balloons) placed in 2005 and 2006 on the 32~m high US Tower (800~m from the Concordia buildings). But the accumulation of ice on the sensors spoiled the results (and eventually broke the sensors). This approch was abandoned to the benefit of fast sonic anemometers able to estimate the structure constant of the temperature $C_n^2$~\citep{2008SPIE.7012E.147T}.

\begin{figure}[ht!]
 \centering
 \includegraphics[width=0.7\textwidth,clip]{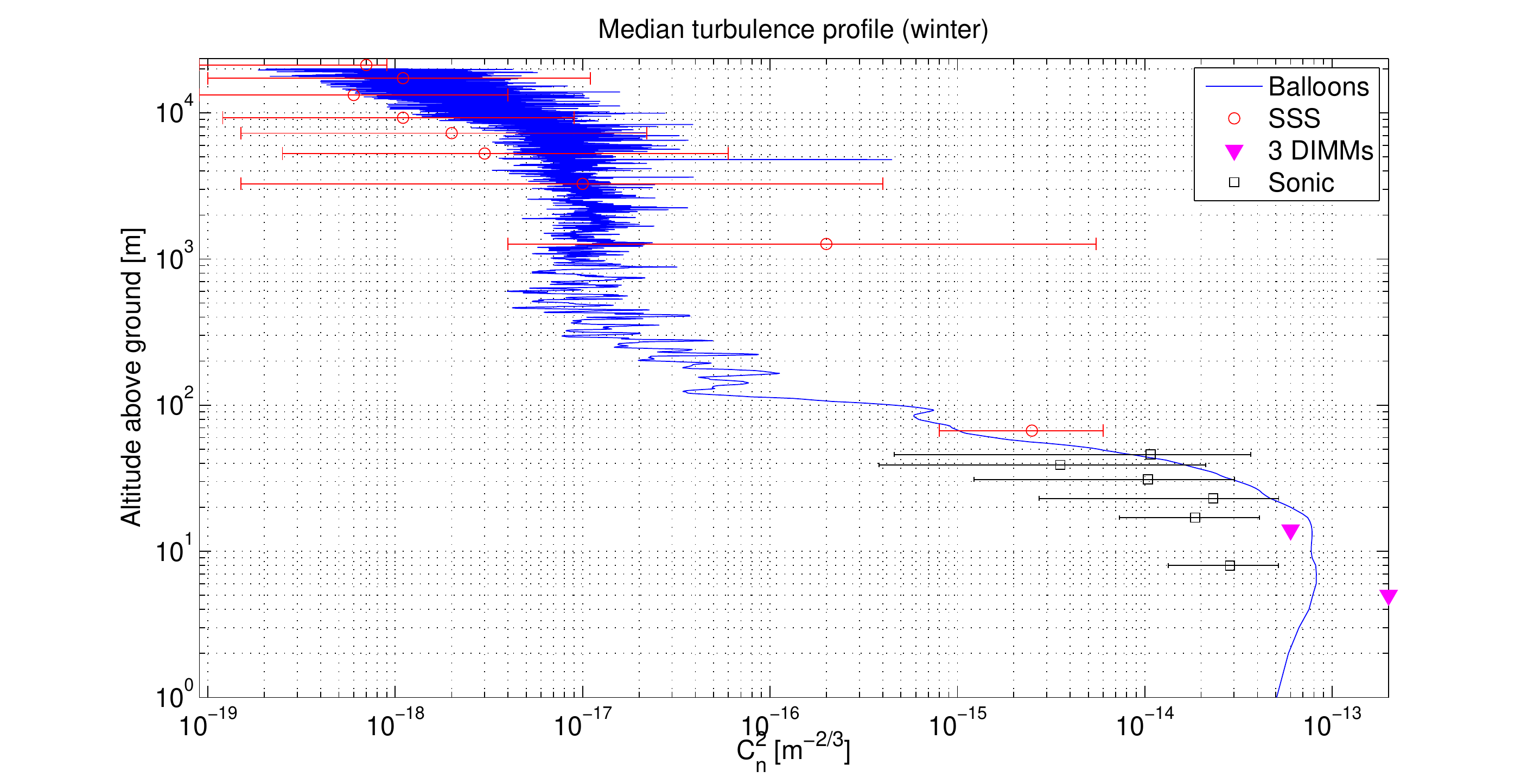}      
  \caption{Median vertical profile of the structure constant $C_n^2$ as a function of the altitude above the ground, measured by various instruments.}
  \label{aristidi:fig2}
\end{figure}

\begin{figure}[ht!]
 \centering
 \includegraphics[width=0.6\textwidth,clip]{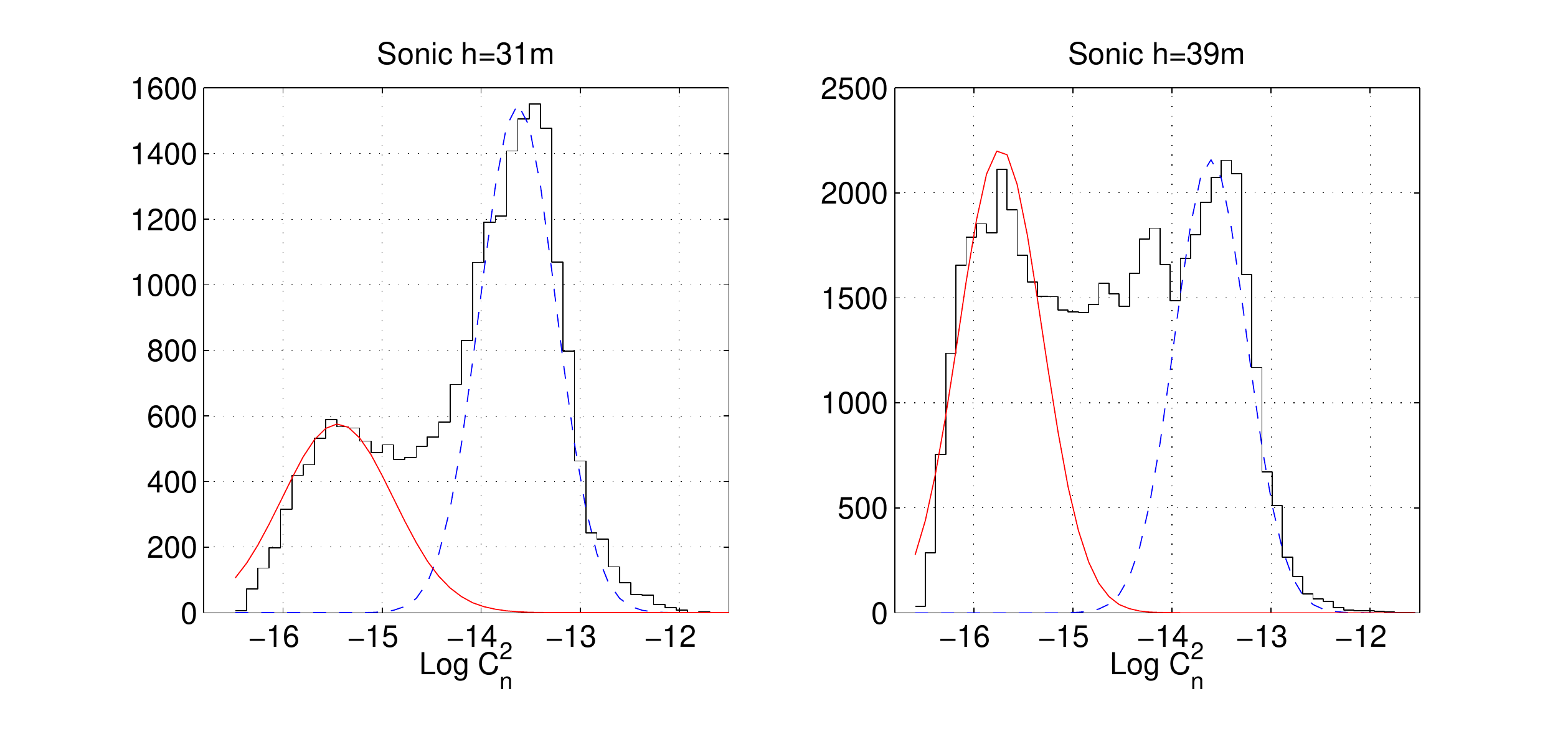}      
 \includegraphics[width=0.3\textwidth,clip]{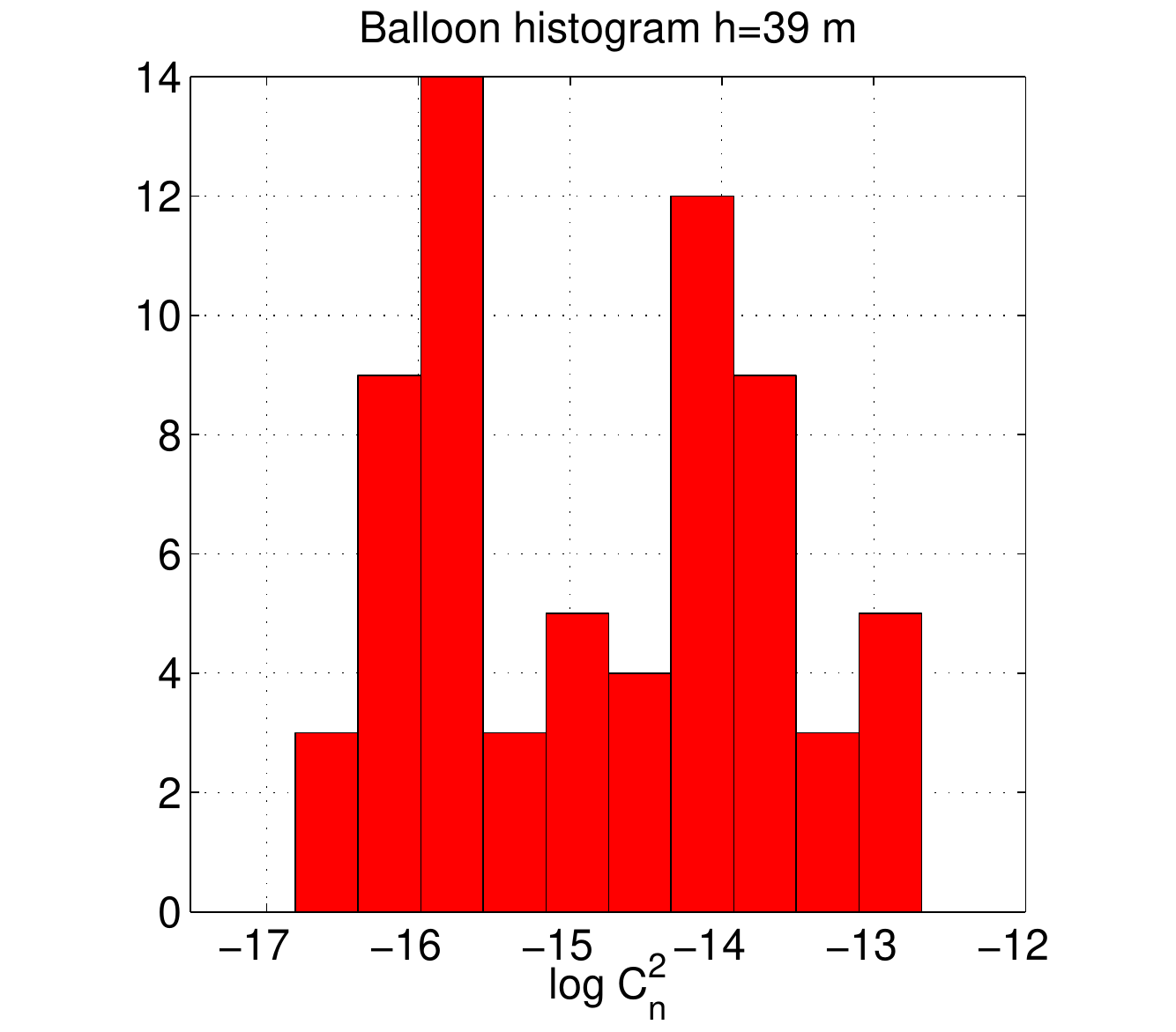}      
  \caption{{\bf Left:} Histogram of Log$(C_n^2)$ in winter (April to September, 2008 to 2011) measured by the sonic anemometer at an elevation of 31~m. The structure constant $C_n^2$ is in m$^{-2/3}$. The data being still under processing, these results are preliminar. {\bf Middle:} Same for the sonic at 39~m. {\bf Right:} Histogram of  Log$(C_n^2)$ measured by the balloon-borne experiment in 2005 at the elevation $h=39$~m ($\simeq$ 80 values. available).}
  \label{aristidi:fig3}
\end{figure}

6 anemometers were deployed on the tower (which was prolongated up to 46~m in 2008) at regular altitudes from 8~m to 45~m. These instruments are periodically heated to prevent frost deposit. The period of the cycle was adjusted empiricaly by the winterers on site to optimize the quantity of usable data. The data are still under processing, but we start to obtain reliable measures as illustrated by the vertical profile of $C_n^2$ in Fig.~\ref{aristidi:fig2}. It is interesting to note that the histograms of all sonic data exhibit a two bumps structure, exactly like the DIMM. Fig.~\ref{aristidi:fig3} shows the histograms of Log$(C_n^2)$ in winter (April to September) for the anemometers at elevations 31~m and 39~m. On the same figure we displayed the histogram of balloon measurements taken in 2005 at the same altitude, which also show the same structure. The abcissa of the center of these bumps is the same in both cases. The leftern bump correspond to situations where the sonic is outside the surface layer and then to the histogram of Log$(C_n^2)$ in the free atmosphere. From all sonic data, we can have an estimation of the median thickness of the SL at a value $h_{SL}=35$~m. Once again, this is a preliminar number but is in agreement with the thickness published by~\citep{2008PASP..120..203T}. Table~\ref{aristidi:table4} summarizes published values of the surface layer thickness at Dome C.

\begin{table}[ht!]
 \centering
 \includegraphics[width=0.8\textwidth,clip]{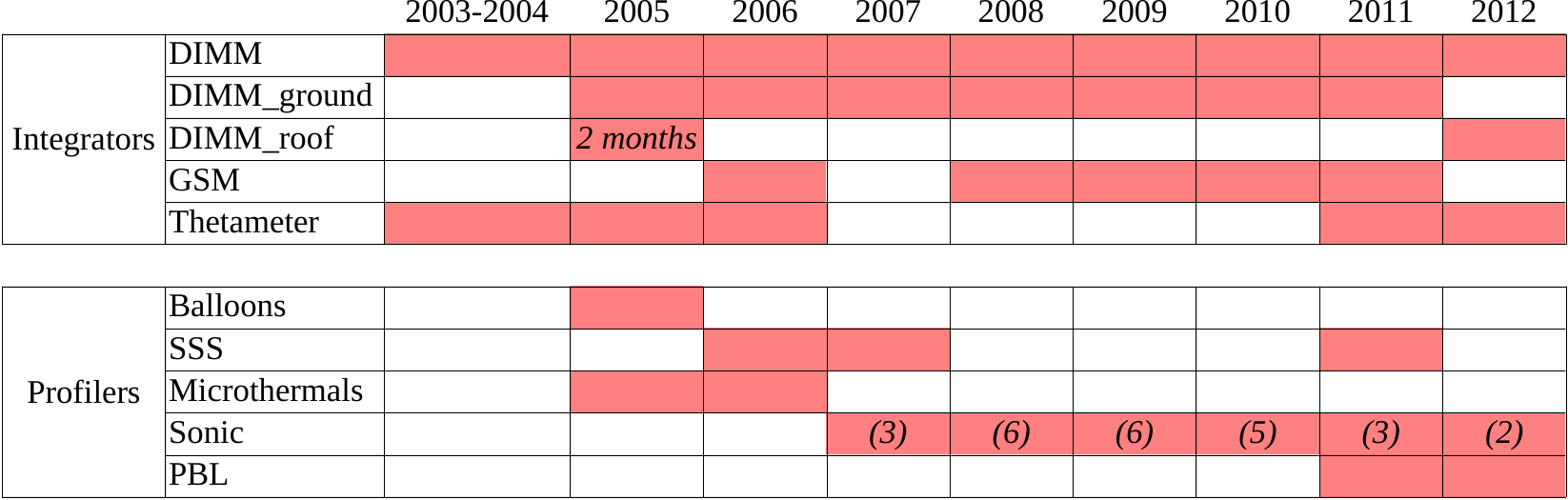}      
  \caption{Operating periods for each instrument (colored boxes). For Sonic we give also the number of available anemometers.}
  \label{aristidi:table1}
\end{table}

%
%
\begin{table}
 \centering
\begin{tabular}{l|c|c|c|c}
                              & Seeing   & Isoplan. angle & Coh. time & Outer scale\\
                              & [arcsec] &   [arcsec]     & [ms]      & [m]\\ \hline
DIMM/GSM                      & 1.7 [1.0 -- 2.4]     &   4.1 [2.7 -- 5.9]         &           & 7.5 [5 -- 11]\\
SSS \citep{2012PASP..124..494G} & 1.0    &   6.9          & 3.4       & \\
Balloons \citep{2008PASP..120..203T}& 1.4&   2.7          & 5.7       & \\
Meso-NH model \citep{2011MNRAS.411..693L} & 1.7  &                &           & \\ \hline
South Pole \citep{1999AAS..134..161M}    & 1.8      &   3.2          & 1.6       &\\
Mauna Kea \citep{1995SPIE.2534..248R}    & 0.6      &  1.9           & 2.7       & 17\\
Paranal  \citep{sarazin_ESO}   & 0.8      &  2.6           & 3.3       & 22\\ \hline
\end{tabular}     
  \caption{Integrated parameters in winter (April to September) at an altitude of 8~m from various instruments. Meso-NH is a computer-based model developped by the Arcetri Observatory (Italy). A comparison with values at other sites is provided (the outer scale value for Mauka Kea is from \citep{schock2002}).}
  \label{aristidi:table2}
\end{table}

%
%
\begin{table}
 \centering
\begin{tabular}{l|c|c|c}
                              & Seeing   & Isoplan. angle & Coh. time \\
                              & [arcsec] &   [arcsec]     & [ms]      \\ \hline
DIMM/GSM \citep{2009AA...499..955A}                   & 0.4      &   4.1          &           \\
SSS \citep{2012PASP..124..494G} & 0.3    &   6.9          & 10.2      \\
Balloons \citep{2008PASP..120..203T}& 0.4&   2.7          & 6.8       \\
AASTINO \citep{2004Natur.431..278L}& 0.3&   5.7          & 7.9       \\
Meso-NH model \citep{2011MNRAS.411..693L} & 0.3  &                &    \\ \hline
\end{tabular}     
  \caption{Free astmosphere integrated parameters in winter (April to September) above the surface layer. AASTINO was an experiment deployed in 2004 by the University of New South Wales, Australia.}
  \label{aristidi:table3}
\end{table}

%
%
\begin{table}
 \centering
\begin{tabular}{l|c}
			 & $h_{SL}$ [$m$]   \\ \hline
Sonic 			 & 35  \\
Balloons \citep{2008PASP..120..203T}& 33 [23 -- 42]  \\
DIMMs \citep{2009AA...499..955A}& 27 [16 -- 54]  \\
Meso-NH model \citep{2011MNRAS.411..693L} & 44 [20 -- 69] \\ \hline
\end{tabular}     
  \caption{Median thickness of the turbulent surface layer, estimated from various instrument as well as the Meso-NH code. Values in brackets are the 1st and 3rd quartile.}
  \label{aristidi:table4}
\end{table}

\section{Conclusions}
We have presented statistics of the optical turbulence at Dome C in winter, using all data collected by our instruments since 2004. We confirm the general scheme of a thin but strong turbulent layer above the surface, which cause the ground seeing to be very poor, as well as the coherence time. Above this surface layer whose thickness is measured between 27m and 35m, the conditions are exceptionnal, the seeing can attain values of 0.4~arcsec half of the time. We also discovered that the distribution of the values of  $C_n^2$ and the seeing display a bimodal structure. This indicates, as pointed by \citep{2009AA...499..955A}, that the upper edge of the surface layer is very sharp and that a telescope at a given altitude would be either inside or outside this surface layer with very little intermediate cases.

\section*{acknowledgements}
The authors gratefully acknowledge the polar agencies IPEV and PNRA, the US NSF and the french agencies INSU and ANR for logistical support and funding. We are in debt to the Dome C local staff and winter-overs of 2005 to 2012 for their assistance.


\bibliographystyle{aa}  
\bibliography{aristidi} 

\end{document}